\documentclass[prb,aps,twocolumn,showpacs]{revtex4}% Physical Review B

\usepackage{graphicx}
\usepackage{latexsym}
\usepackage{amsmath}
\usepackage{bm}
\usepackage{textcase}

\newcommand{\be}{\begin{eqnarray}}
\newcommand{\ee}{\end{eqnarray}}

\begin{document}

\title{Superfluidity and solid orders in two-component Bose gas with dipolar 
interactions \\
in an optical lattice}

\date{\today}

\author{Yoshihito Kuno} 
\author{Keita Suzuki} 
\author{Ikuo Ichinose}
\affiliation{%
Department of Applied Physics, Nagoya Institute of Technology,
Nagoya, 466-8555 Japan}
%{\today}

\begin{abstract}
In this paper, we study an extended bosonic t-J model in an
optical lattice, which describes two-component hard-core bosons with 
nearest-neighbor (NN) pseudo-spin interactions, and also inter- and intra-species
dipole-dipole interactions (DDI).
In particular, we focus on the case in which two component hard-core bosons have 
anti-parallel polarized dipoles with each other.
The global phase diagram is studied by means of the Gutzwiller variational method 
and also the quantum Monte-Carlo simulations (QMC). 
The both calculations show that a stripe solid order, besides a checkerboard one,
appears as a result of the DDI.
By the QMC, we find that two kinds of supersolid (SS) form, i.e.,
checkerboard SS and stripe SS, and we also verify the existence of some exotic phase
between the stripe solid and checkerboard SS.
Finally by the QMC, we study the t-J-like model, which was experimentally realized 
recently by A. de Paz et al. [Phys. Rev. Lett. {\bf 111}, 185305 (2013)].
\end{abstract}


\maketitle
%%%%%%%%%%%%%%%%%%%%%%%%%%%%%%
\section{Introduction}

In recent years, cold atomic systems play a very important role for study
of the condensed matter physics.
In particular, the cold atomic system in an optical lattice (OL) \cite{optical} is
sometimes regarded as a feasible simulator searching for new type of quantum states.
Cold atomic systems in an OL are versatile and effects of defect and 
impurity are negligibly small.
By the quantum simulators, some important subjects have been studied 
including; the strongly-correlated systems\cite{Santos}, lattice gauge 
theory\cite{LGT} and cosmology \cite{Cosmological}, etc.
  
From the above point of view, we are interested in the exotic quantum state
in the cold atomic gases called a supersolid (SS)\cite{SS1}, which has both 
a crystalline and a superfluid (SF) orders.  
While many interesting works on this subject\cite{Batrouni1,SS} have
been reported for single-component cold atomic gases, the study on 
two-component boson systems is still inadequate. 
In the present paper, we shall study the bosonic 
t-J model (B-t-J)\cite{BtJ0,BtJ,BtJ1,Altman} 
with dipole-dipole interactions (DDI)\cite{DDI}, as 
the long-range nature of the DDI possibly generates interesting phases
including the SS. 

To study the phase diagram in detail, we shall employ both the Gutzwiller variational 
method and numerical quantum Monte-Carlo simulations (QMC). 
To perform the QMC, we use the effective field-theory model of the B-t-J model
derived in the previous paper\cite{KKI2}.
All relevant quantum fluctuations are included in the QMC of the effective model. 
The phase diagrams obtained by the above two methods are compared with 
each other and effects of the quantum fluctuations are discussed.
  
This paper is organized as follows.
In the first subsection of Sec.II, we shall introduce the B-t-J model and 
briefly explain the derivation of the effective field theory. 
In the derivation, the hard-core constraint of the B-t-J model is faithfully treated
by using the slave-particle representation.
In the second subsection of Sec.II, we consider the DDI and introduce its effects
into the B-t-J model. 
In the present paper, we consider the case in which dipoles of two-component boson 
are anti-parallel with each other.
In this case, the DDI are nothing but the $z$-component pseudo-spin interactions.
Then we call the resultant mode extended B-t-J model.
In Sec.III, we study the phase diagram by using the Gutzwiller method, which
is a kind of the mean-field approximation.
In Sec.IV, the results obtained by means of the QMC are shown and discussed.
The detailed investigation of the global phase diagram is given, in particular,
states in the region of the competing orders of the solid and SF are
discussed in detail.
In Sec.V, we introduce and study an anisotropic B-t-J model (called 
B-t-J-like model), which was recently realized by the experiment\cite{Santos}.
By the QMC, we show that the experimentally observation is reproduced 
in the model.
Section VI is devoted for conclusion.

%%%%%%%%%%%%%%%%%%%%%%%%%%%%%%
\setcounter{equation}{0}
\section{Bosonic $\mbox{t-J}$ Model with DDI and derivation of effective model}

\subsection{Bosonic $\mbox{t-J}$ model and the  slave-particle representation}

System of two-species Bose gas in an optical lattice with the strong
on-site repulsions is often described by the B-t-J model.
Its relationship to the Bose-Hubbard model was discussed in the 
previous papers.
In the present paper, we regard the B-t-J model is a canonical model
for the strong on-site repulsive Bose-gas system.
Hamiltonian of the B-t-J model is given as follows\cite{BtJ0,BtJ,BtJ1,Altman,SpinM},
\begin{eqnarray}
H_{\rm tJ}&=&-\sum_{\langle i,j\rangle} t(a^\dagger_{i}a_j
+b^\dagger_{i}b_j+\mbox{h.c.})
+J_z\sum_{\langle i,j\rangle}S^z_{i}S^z_j  \nonumber  \\
&& -J_{\rm XY}\sum_{\langle i,j\rangle}(S^x_{i}S^x_j+S^y_{i}S^y_j)
%-{\mu}_c\sum_{r}(1-n_{ar}-n_{br}),
\label{HtJ}
\end{eqnarray}
where $a^\dagger_i$ and $b^\dagger_i$ are 
boson creation operators\cite{HCboson} at site $i$,
pseudo-spin operator is given as $\vec{S}_i={1 \over 2}B^\dagger_i\vec{\sigma}B_i$ with
$B_i=(a_i,b_i)^t$ and the Pauli spin matrices $\vec{\sigma}$,
and  $\langle i,j\rangle$ stands for nearest-neighbor (NN) sites of the lattice.
We shall consider two dimensional square lattice in the following study.
The first $t$-term of the Hamiltonian (\ref{HtJ}) is the hopping term of the
$a$ and $b$-atoms, the second $J_z$-term represents the interaction
between atoms at the NN sites, and the third $J_{\rm XY}$-term 
enhances the coherence of the relative phase 
of the $a$ and $b$-atomic fields.
From the definition of the Pauli matrix $\sigma_z$, it is obvious that 
$J_z$-term corresponds to a repulsive intra-species interaction and an attractive 
inter-species interaction for $J_z>0$.

Physical Hilbert space of the B-t-J model consists of states in which
the total particle number at each site is strictly restricted to be less than unity.
In order to incorporate this local constraint 
faithfully, we use the following slave-particle representation\cite{BtJ,BtJ1},
\begin{eqnarray}
&& a_i=\phi^\dagger_i \varphi_{i1}, \;\;\; 
b_i=\phi^\dagger_i \varphi_{i2},  \label{slave}  \\
&& \Big(\phi^\dagger_i\phi_i+\varphi^\dagger_{i1}\varphi_{i1}+
\varphi^\dagger_{i2}\varphi_{i2}-1\Big)
|\mbox{phys}\rangle =0,
\label{const}
\end{eqnarray}
where $\phi_i$ is a boson operator that {\em annihilates hole} at site $i$,
whereas $\varphi_{1i}$ and $\varphi_{2i}$ are bosons that represent the pseudo-spin 
degrees of freedom.
$|\mbox{phys}\rangle$ is the physical state of the slave-particle Hilbert space. 

The previous numerical study of the B-t-J model\cite{BtJ,BtJ1,KKI} 
showed that there appear various phases including super-fluid (SF) with 
Bose-Einstein condensation (BEC), state with the pseudo-spin long-range order, 
etc.
For the most of the parameter regions, the QMC show that 
the density fluctuation of particles at each lattice site is stable even 
in the spatially inhomogeneous states like a phase-separated state.
From this observation, we expect that there appears the following term effectively,
\begin{eqnarray}
H_{\rm V} &=& {V_0 \over 4}\sum_i\Big(
(\varphi^\dagger_{1i}\varphi_{1i}-\rho_{1i})^2+
(\varphi^\dagger_{2i}\varphi_{2i}-\rho_{2i})^2 \nonumber \\
&& \hspace{2cm} +(\phi^\dagger_{i}\phi_i-\rho_{3i})^2
\Big),
\label{HV}
\end{eqnarray}
where $\rho_{1i}$ etc are the parameter that controls the densities of
$a$-atom and $b$-atom at site $i$, and $V_0(>0)$ controls their
fluctuations around the mean values.
It should be remarked here that the expectation value of the particle
numbers in the physical state $|\mbox{phys}\rangle$ are given as
$\langle a^\dagger_i a_i \rangle\equiv
\mbox{Tr}_{\rm phys}(a^\dagger_ia_i)
=\mbox{Tr}_{\rm phys}(\varphi^\dagger_{1i}\varphi_{1i})$
and similarly 
$\langle b^\dagger_i b_i \rangle=
\mbox{Tr}_{\rm phys}(\varphi^\dagger_{2i}\varphi_{2i})$,
where $\mbox{Tr}_{\rm phys}$
denotes the trace over the states satisfying the local 
constraint (\ref{const}).
Therefore
the constraint (\ref{const}) requires $\sum_{\sigma=1}^3\rho_{\sigma i}=1$ at each
site $i$.
The values of $V_0$ and $\rho_{\sigma i} (\sigma=1,2,3)$ are to be determined 
in principle by $t, \ J_z, \ J_{\rm XY}$ and 
filling factor, but here we add $H_{\rm V}$ to $H_{\rm tJ}$ by hand
and regard the parameter $V_0$ in  $H_{\rm V}$ as a free parameter,
whereas $\rho_{\sigma i} (\sigma=1,2,3)$ are to be determined accurately
by $H_{\rm tJ}$.
In other words, we take the extended B-t-J model $H_{\rm tJ}+H_{\rm V}$
as a canonical model and regard $H_{\rm V}$ as a residual one-site repulsion that 
cannot be incorporated by the hard-core constraint. 
However, we expect that the original B-t-J model has a similar 
phase diagram to that of the extended B-t-J model.
See later remarks on this point.

By means of the path-integral method, the partition function $Z$ is expressed
as follows by introducing the imaginary time $\tau$,
\begin{eqnarray}
Z&=& \int [D\phi D\varphi_1 D\varphi_2]
\exp\Big[-\int d\tau\Big(\bar{\varphi}_{1i}(\tau)\partial_\tau \varphi_{1i}(\tau)
\nonumber \\
&& \hspace{1cm} +\bar{\varphi}_{2i}(\tau)\partial_\tau \varphi_{2i}(\tau)
 +\bar{\phi}_i(\tau)\partial_\tau \phi_i(\tau)  \nonumber \\
&& \hspace{1cm} +H_{\rm tJ}+H_{\rm V}
\Big)\Big],
\label{Z}
\end{eqnarray}
where $H_{\rm tJ}$ and $H_{\rm V}$ are expressed by the slave particles (\ref{slave})
and the above path integral is evaluated under the constraint (\ref{const}).
The direct QMC is not applicable to the system (\ref{Z}) due to the Berry phases
$(\bar{\varphi}\partial_\tau \varphi)$ etc, and therefore
we separate the path-integral variables 
$\varphi$'s and $\phi$ as 
\begin{eqnarray}
&&\varphi_{1i}=\sqrt{\rho_{1i}+\ell_{1i}}\exp(i\omega_{1i}), \nonumber \\  
&&\varphi_{2i}=\sqrt{\rho_{2i}+\ell_{2i}}\exp(i\omega_{2i}), \label{param1} \\
&&\phi_i=\sqrt{\rho_{3i}+\ell_{3i}}\exp(i\omega_{3i}), \nonumber
\end{eqnarray}
and then integrate out the (fluctuation of) the radial degrees of freedom,
$\ell_{\sigma i} \ (\sigma=1,2,3)$.
By the existence of the term $H_{\rm V}$, the integration can be performed
straightforwardly.
There exists a constraint like 
$\ell_{1i}+\ell_{2i}+\ell_{3i}=0$
on performing the path-integral over the radial degrees of freedom.
But this constraint can be readily incorporated by using a Lagrange 
multiplier $\lambda_i(\tau)$,
$$
\prod_\tau\delta(\ell_{1i}+\ell_{2i}+\ell_{3i})=\int d\lambda_i
e^{i\int d\tau (\ell_{1i}+\ell_{2i}+\ell_{3i})\lambda_i}.
$$
The variables $\ell_{\sigma i} \ (\sigma=1,2,3)$ also appear in $H_{\rm tJ}$,
but we ignore them by simply replacing 
$\varphi_{\sigma i} \rightarrow \sqrt{\rho_{\sigma i}}\exp(i\omega_{\sigma i})$,
and then we have
\begin{eqnarray}
&&\int d\lambda_i d\ell_{i}e^{\int d\tau\sum_{\sigma=1}^3(-V_0(\ell_{\sigma,i})^2
+i\ell_{\sigma,i}(\partial_\tau \omega_{\sigma,i}+\lambda_i))} \nonumber   \\
&& \hspace{1cm} =\int d\lambda_i e^{-{1 \over 4V_0}\int d\tau\sum_\sigma
(\partial_\tau \omega_{\sigma,i}+\lambda_i)^2},
\label{integral}
\end{eqnarray}
where we have ignored the terms like $\int d\tau \partial_\tau \omega_{\sigma, i}$
by the periodic boundary condition for the imaginary time.
The resultant quantity on the RHS of (\ref{integral}) is positive definite,
and therefore the numerical study by the QMC can be performed
without any difficulty.
It should be remarked that the Lagrange multiplier $\lambda_i$
in Eq.(\ref{integral}) behaves as a gauge field, i.e.,
the RHS of (\ref{integral}) is invariant under the following ``gauge transformation",
$\omega_{\sigma,i}\rightarrow \omega_{\sigma,i}+\alpha_i, \ 
\lambda_i \rightarrow \lambda_i-\partial_\tau \alpha_i$.
In the practical calculation, we shall show that all physical quantities are
invariant under the above gauge transformation.

Here, remarks are in order.
\begin{enumerate}
\item 
The direct QMC of the system $Z$ in Eq.(\ref{Z})
is impossible for the Berry phases $\bar{\varphi}\partial_\tau \varphi$
are pure imaginary.
However by the integrating over the density fluctuations $\ell_{\sigma i}$,
the action becomes positive-definite as Eq.(\ref{integral}) shows,
and then the QMC is applied without any difficulty.
\item
In order to integrate over $\ell_{\sigma i}$, we have introduced the density-fluctuation
term $H_{\rm V}$.
Effectively similar terms to $H_{\rm V}$ are generated from the terms in 
the Hamiltonian of the original B-t-J model (\ref{HtJ}).
For example, a rough estimation for the density fluctuation of the $a$-atom $\delta\rho_{ai}$ gives,
\begin{equation}
\Big(t{{\cal P} \over {\rho}_{ai}}+J_{\rm XY}\sqrt{{\rho}_{ai}}
{{\cal Q} \over (\sqrt{{\rho}_{bi}})^3}\Big)(\delta\rho_{ai})^2,
\label{delrho}
\end{equation}
where ${\rho}_{ai}$ (${\rho}_{bi}$) is the mean value of the $a$-atom
($b$-atom) density, and the positive parameters ${\cal P}$ and ${\cal Q}$ are determined
by the NN correlations of the phase degrees of freedom of the atomic fields like
${\cal P}=\langle \cos (\theta_{ai}-\theta_{aj})\rangle$ where $\theta_{ai}$
is the phase of the $a$-atom field.
In the QMC in Sec.IV, we fix the values of $t$ and
$J_{\rm XY}$, and the values of ${\rho}_{ai}$ and ${\rho}_{bi}$ are determined
by the B-t-J model (\ref{HtJ}) quite accurately.
It is difficult to obtain the coefficient in Eq.(\ref{delrho}) accurately, but it is expected
that coefficient of Eq.(\ref{delrho}) is fairly stable against the variations of
${\rho}_{ai}$ and ${\rho}_{bi}$ as we require the constraint 
${\rho}_{ai}+{\rho}_{bi}=1-$(constant hole density) in the calculation and also by
the behavior of the correlators ${\cal P}$ and ${\cal Q}$.
Furthermore for the system $H_{\rm tJ}+H_{\rm V}$, we have verified 
by the practical calculation
that a change of value of $V_0$ in $H_{\rm V}$ does not substantially influence 
the global phase diagram of the system,
although the SF region slightly increases for smaller $V_0$ as larger density fluctuation
stabilizes the phase degrees of freedom by the density-phase uncertainty principle.
See Ref.\cite{KKI2}, in particular, the left panels of figure 1.
Then it is naturally expected that the obtained phase diagram of the constant 
$V_0$-system by the QMC faithfully describes the phase structure of 
the original B-t-J model as well as the extended B-t-J model with the $H_{\rm V}$-term.
\item
The partition function obtained by performing the integral in Eq.(\ref{integral}) 
depends on the local density of the bosons $\rho_{\sigma i}$.
We treat the density difference
$\Delta\rho_i\equiv\rho_{1i}-\rho_{2i}=\rho_{ai}-\rho_{bi}$ 
as a variational parameter {\em while keeping one-site hole density fixed}, i.e.,
$\rho_{1i}+\rho_{2i}=\rho_{ai}+\rho_{bi}=$constant.
These treatments obviously preclude the possibility a phase separated state. 
The previous study by means of a Gross-Pitaevskii equation and QMC\cite{KSI}
shows that such a phase separated state does not appear in the B-t-J model
(\ref{HtJ}). 
Therefore this treatment is justified.
\item 
The last remark concerns the Hamiltonian (\ref{HtJ}) itself.
Originally, the B-t-J model was derived as an effective model of the Bose-Hubbard
model in the large on-site-repulsion limit.
By integrating out the multiple-particle states at each site, the NN terms of
the pseudo-spin interactions appear.
At present, however, the interactions between atoms located at the NN sites
can be generated and their strength is controlled by using the DDI.
Then the Hamiltonian (\ref{HtJ}) can be regarded as an original Hamiltonian,
and it is quite natural to add the Hubbard term $H_{\rm V}$ in Eq.(\ref{HV}) to
$H_{\rm tJ}$.
In this case, the $a$ and $b$-atoms are not a hard-core boson and their density
can take arbitrary values.
Effective model of the system is derived by a similar method to the above, but
the use of the slave-particle representation is not needed.
\end{enumerate}

As we explained in the introduction, we shall study Bose gases with the DDI in this paper.
In Sec.IIB, we briefly explain the DDI, which gives a long-range interaction
between the $z$-component of the pseudo-spin $S^z$.

%%%%%%%%%%%%%%%%%%%%%%%%%%%%%%%%%%%

\subsection{Realization of long-range spin interactions via the DDI} 

\begin{figure}[ht]
\begin{center}
\includegraphics[width=7cm]{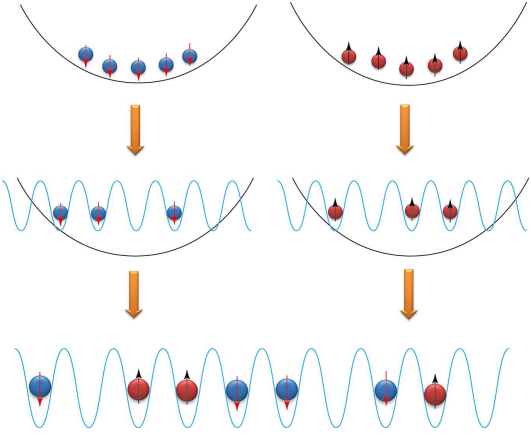}
\caption{(Color online)
Experimental setup for two component dipolar bosons in an OL.
The dipole-dipole interactions generate the long-range repulsive and attractive
interactions between two kinds of bosons.}
\label{DDIsk}
\end{center}
\end{figure}

%As we explained in the previous sections, we consider the Bose gass of the DDI. 
When the atoms have a magnetic or electric dipole, terms describing
the DDI \cite{DDI} should be added to the B-t-J model Hamiltonian.
We first consider some specific case in which the $a$-atom has the upward dipole,
whereas the $b$-atom has the downward one perpendicular to the OL.
See Fig.\ref{DDIsk}.
We regard this system as a canonical system and clarify its phase diagram
in the subsequent sections.
The system of the strongly-correlated dipolar gas, which was realized 
by the experiments recently, will be considered in Sec.VI, because it has rather
strong anisotropy in couplings.

The DDI is generally given as
\begin{eqnarray}
\hat{H}_{d}=d^2\frac{\hat{\bf S}_{1}\cdot \hat{\bf S}_{2}-3(\hat{\bf S}_{1}\cdot \hat{\bf r})(\hat{\bf S}_{2}\cdot \hat{\bf r})}{4\pi r^{3}},
\label{Hd}
\end{eqnarray}	
where $d^2=\mu_{0}(g\mu_{B})^2$ ($\mu_0$ being the magnetic permeability of vacuum,
$g$ the Lande factor, and $\mu_B$ the Bohr magneton), 
$\hat{\bf S}_j$ ($j=1,2$) is dipole-moment vector
of the $j$-th atom and $\hat{\bf r}=\frac{{\bf r}}{r}$
with the relative position vector ${\bf r}$ of the atoms. 
In the present canonical system, $\hat{\bf S}_1 // \hat{\bf S}_2$ and 
$\hat{\bf S}_1, \hat{\bf S}_2\perp {\bf r}$, and therefore the only the first term
on the RHS of Eq.(\ref{Hd}) contributes.
In Fig.\ref{DDIsk}, an experimental manipulation for realizing the canonical system
is schematically shown; 
First we prepare independently $a$-boson with up-polarized state and 
$b$-boson with down-polarized state in two magnetic traps. 
Second, the OL is created in each trap. 
Finally, we combine these two systems quasi-statically and lower the
temperature. 
As a result of the strong repulsions between atoms and the finite
hopping amplitude, the total particle number at each site of the OL is
less than unity.
Furthermore, due to the angular-momentum conservation, direction of the dipole
does not change under the hopping of atoms.

%%%%%%%%%%%%%%%%%%%%%%%%%%%%%%%%%%%%%%%%%%%%%%%%%%%%%%%%%%%
%FIG.1
%%%%%%%%%%%%%%%%%%%%%%%%%%%%%%%%%%%%%%%%%%%%%%%%%%%%%%%%%%%

Though the DDI has long-range nature, we shall consider only the
NN coupling and the next-nearest-neighbor (NNN) coupling in the OL.      
In the present dipole configuration, the DDI reduces an inter-species attraction 
and intra-species repulsion. 
Then $\hat{H}_d$ in Eq.(\ref{Hd}) effectively generates the following terms
\begin{eqnarray}
V_{\rm DDI}&=&\sum_{\langle i,j\rangle}
V_{\rm NN}(n_{ai}n_{aj}+n_{bi}n_{bj}-n_{ai}n_{bj}-n_{bi}n_{aj})\nonumber\\
       &+&\sum_{\langle\langle i,j\rangle\rangle}
V_{\rm NNN}(n_{ai}n_{aj}+n_{bi}n_{bj}-n_{ai}n_{bj}-n_{bi}n_{aj})\nonumber\\
       &=&\sum_{\langle i,j\rangle}
V_{\rm NN}S^{z}_{i}S^{z}_{j}+
\sum_{\langle\langle i,j\rangle\rangle}V_{\rm NNN}S^{z}_{i}S^{z}_{j},
\label{VDDI}
\end{eqnarray}
where $n_{ai}=a^\dagger_{i}a_{i}$ etc, and
we have used $S^z_i={1 \over 2}(n_{ai}-n_{bi})$, and $\langle\langle i,j\rangle\rangle$
stands for NNN sites.
The parameters
$V_{\rm NN}$ and $V_{\rm NNN}$ are given by the overlap integral
of the lowest level Wannier functions (s-wave) on the OL sites.
The $V_{\rm DDI}$ term in Eq.(\ref{VDDI}) is to be added to the Hamiltonian of 
the B-t-J model. 
In the following studies, we shall consider the system described by 
$H_{\cal T}\equiv H_{\rm tJ}+V_{\rm DDI}$. 

%%%%%%%%%%%%%%%%%%%%%%%%%%%%%%%%%%%%%%%%%%%%%%%%%%%%%%%%%%%%%%%%%

\setcounter{equation}{0}
\section{Gutzwiller variational method}

Mean-field theory (MFT) is widely used to study phase diagram
of condensed matter systems.
In this section, we employ the Gutzwiller variational method, 
which is a kind of the MFT, to investigate the phase diagram of the
system $H_{\cal T}$ at vanishing temperature ($T$).
From the results of the previous studies\cite{SS}, we expect the appearance of 
two kinds of solid order, i.e., checkerboard-solid (CBSo) and stripe-solid (SSo) orders 
in certain parameter regions.
The solid order is a spatial pattern of the atomic densities
and is noting but a pseudo-spin order in the B-t-J model as $S^z_i$ is given by
$S^z_i={1 \over 2}(n_{ai}-n_{bi})$. 

As the present model describes the strong on-site repulsion limit, the
physical state at each site consists of the following three state: 
$|a\rangle$ (single $a$-boson), $|b\rangle$ (single $b$-boson) and 
$|0\rangle$ (empty=hole). 
By using the above three basis vectors, we construct a variational wave function 
corresponding to the state with the double SF (2SF) and/or the CBSo,
\begin{eqnarray}
&&|\Phi_{\rm 2SF-CB}\rangle \nonumber\\
&&=\Pi_{i\in A}\biggl[\sin\frac{\theta_{i}}{2}\biggl(\sin\frac{\chi_{i}}{2}
a^{\dagger}_{i}
+\cos\frac{\chi_{i}}{2}b^{\dagger}_{i}\biggr)+\cos\frac{\theta_{i}}{2}\biggr]|0\rangle\nonumber\\
&&\times\Pi_{i\in B}\biggl[\sin\frac{\theta_{i}}{2}\biggl(\cos\frac{\chi_{i}}{2}
a^{\dagger}_{i}
+\sin\frac{\chi_{i}}{2}b^{\dagger}_{i}\biggr)+\cos\frac{\theta_{i}}{2}\biggr]|0\rangle , 
\nonumber\\
\label{WF1}
\end{eqnarray}
where the label A(B) stands for the even (odd) sub-lattice, and the parameters
$(\theta_i,\chi_i)$ are to be determined by the variational method.
In the MFT level, we reduce the local variables $\theta_{i}$ and $\chi_{i}$ to global ones,
$(\theta_A, \theta_B)$ and $(\chi_A,\chi_B)$. 
It should be noticed that the state of the wave function (\ref{WF1})
has the discrete translational symmetries of the twice lattice spacing in both the $x$ 
and $y$-directions.

Another possible solid order is the SSo.
Variational wave function that describes the SSo and 2SF is given as
\begin{eqnarray}
&&|\Phi_{\rm 2SF-SSo}\rangle = \nonumber\\
&&\Pi_{i\in x_{o}}\biggl[\sin\frac{\theta_{i}}{2}\biggl(\sin\frac{\chi_{i}}{2}
a^{\dagger}_{i}
+\cos\frac{\chi_{i}}{2}b^{\dagger}_{i}\biggr)+\cos\frac{\theta_{i}}{2}\biggr]|0\rangle\nonumber\\
&\times&\Pi_{i\in x_{e}}\biggl[\sin\frac{\theta_{i}}{2}\biggl(\cos\frac{\chi_{i}}{2}
a^{\dagger}_{i}
+\sin\frac{\chi_{i}}{2}b^{\dagger}_{i}\biggr)+\cos\frac{\theta_{i}}{2}\biggr]|0\rangle, \nonumber\\ 
\label{WF2}
\end{eqnarray}
where the site label $x_{o}$($x_{e}$) denotes odd (even) line sub-lattice
in the $x$-direction corresponding to the stripe pattern. 
The above wave function (\ref{WF2}) has the discrete translational symmetry
of the twice lattice spacing in the $x$-direction and the ordinary one of
the single lattice spacing in the $y$-direction.

From the wave functions (\ref{WF1}) and (\ref{WF2}),
we calculate the expectation value of the the Hamiltonian 
$H_\mu\equiv H_{\cal T}-\mu\sum(a^{\dagger}_{i}a_{i}+b^{\dagger}_{i}b_{i})$,
where $\mu$ is the chemical potential, and obtain
\begin{eqnarray}
{E_{\rm CS}\over J_{\rm XY}N_s}
 &\equiv &\langle \Phi_{\rm 2SF-CB}|H_{\mu}|\Phi_{\rm 2SF-CS}\rangle
/(J_{\rm XY}N_s)\nonumber\\
      &= & -\tilde{t}\sin^{2}\theta\sin\chi+2(-\tilde{J}_{\rm zNN}+\tilde{J}_{\rm zNNN})
\sin^{4}\frac{\theta}{2}\cos^{2}\chi\nonumber\\
      &&-\frac{1}{8}\sin^{4}\theta\sin^{2}\chi - \mu \sin^{2}\frac{\theta}{2},
\label{ECS}\\
{E_{\rm SS}\over J_{\rm XY}N_s}
&\equiv &\langle \Phi_{\rm 2SF-SSo}|H_{\mu}
|\Phi_{\rm 2SF-SSo}\rangle /(J_{\rm XY}N_s)\nonumber\\
      &= & -\frac{1}{2}\tilde{t}\sin^{2}\theta(1+\sin\chi)-2\tilde{J}_{\rm zNNN}
\sin^{4}\frac{\theta}{2}\cos^{2}\chi\nonumber\\
      &&-\frac{1}{8}\sin^{4}\theta\sin^{2}\chi - \mu \sin^{2}\frac{\theta}{2}.
\label{ESS}
\end{eqnarray}
In Eqs.(\ref{ECS}) and (\ref{ESS}), 
the variational energies are normalized by $J_{\rm XY}N_s$, where $N_s$
is the total number of sites, and thus, 
$\tilde{t}\equiv t/J_{\rm XY}$, $\tilde{J}_{\rm zNN}
\equiv J_{\rm zNN}/J_{\rm XY}$ and 
$\tilde{J}_{\rm zNNN}\equiv J_{\rm zNNN}/J_{\rm XY}$, where
 $J_{\rm zNN}\equiv J_{z}+V_{\rm NN}$ and $J_{\rm zNNN}\equiv V_{\rm NNN}$.

\begin{figure}[ht] 
\begin{center} 
\includegraphics[width=7cm]{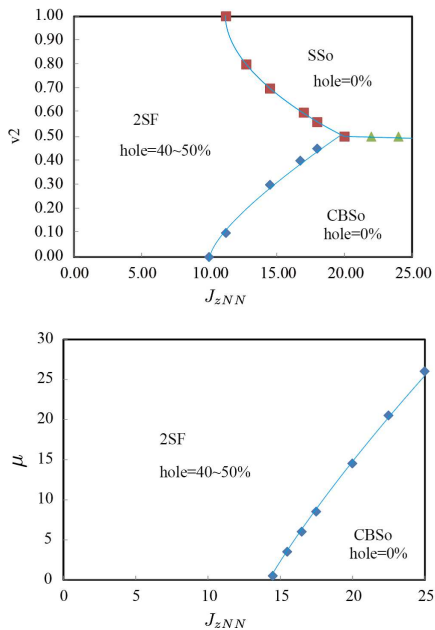}
\caption{(Color online)
Phase diagrams at $T=0$ in the grand-canonical ensemble obtained by the MFT.
The upper panel is $\mu=0$ phase diagram in the $(\tilde{J}_{zNN}-v_2)$-plane,
while the lower panel is the one in the $(\mu$-$\tilde{J}_{zNN})$-plane
for $v_2\equiv J_{\rm zNNN}/J_{\rm zNN}=0.3$. 
In the phase diagrams of the MFT, the SS does not exist.
}
\label{PD1}
\end{center}
\end{figure}

From Eqs.(\ref{ECS}) and (\ref{ESS}), it is rather straightforward to obtain 
the lowest energy states by varying values of $\theta$ and $\chi$, and then
the global phase diagram is obtained. 
The obtained phase diagram is shown in the upper panel in Fig.\ref{PD1},
where $v_2=J_{\rm zNNN}/J_{\rm zNN}$.
The lower panel in Fig.\ref{PD1} is the phase diagram in the
$(\mu$-$\tilde{J}_{\rm zNN})$ plane. 
It is obvious that at the MFT level, the SS does not form and the solid phases, 
the CBSo and SSo, exist only at the vanishing hole density.
On the other hand, the 2SF phase has a finite hole density, 
in particular, the maximal density is 30\%.
The above results are in agreement with the previous results of the MFT 
for the one-component Bose Hubbard model in Refs.\cite{Altman,Batrouni1}, 
which showed that the SS does not form
and a direct phase transition from the CBSo to SSo takes place.

%%%%%%%%%%%%%%%%%%%%%%%%%%%%%%%%%%%%%%%%%%%%%%%%%%%%%%%%%%%
%FIG.2
%%%%%%%%%%%%%%%%%%%%%%%%%%%%%%%%%%%%%%%%%%%%%%%%%%%%%%%%%%%       

\setcounter{equation}{0}
\section{Quantum Monte-Carlo simulation}

%\subsection{Phase diagram}

In this section, we study the extended B-t-J model, $H_{\cal T}+H_{\rm V}$, by
the QMC.
In particular, we are interested in the global phase diagram and the region
in which the SS forms.
As we explained in the introduction, we expect that the SSs with 
different solid orders appear as a result of the DDI. 

For the practical QMC, we put the lattice spacing of the OL, $a_{L}$,
to unit of length. 
We also introduce a discretized lattice for the imaginary-time $\tau$
with the lattice spacing $\Delta\tau$. 
Then, the model is defined on the three-dimensional (3D) space-time lattice, and 
we denote the site of 3D lattice $i,j$, etc hereafter. 

The previous study\cite{KSI} on the B-t-J model $H_{\rm tJ}$ shows that
holes are distributed quite homogeneously except for a very specific 
parameter region in which a phase-separated state forms. 
Therefore, we assume a homogeneous distribution of holes also in the present system 
and the put the hole density at each site to $30\%$, i.e.,
$\rho_{3,i}=\rho_3=0.3$. 
On the other hand, the density difference of the $a$ and $b$-atoms at site $i$,
$\Delta \rho_i=\rho_{1i}-\rho_{2i}$ is a variational variable
and is determined by the maximal free-energy condition.
See later discussion.  

Effective lattice model of the extended B-t-J model $H_{\cal T}+H_{\rm V}$ is derived
from Eqs.(\ref{Z}) and (\ref{integral}).
The partition function and action of the effective model are given as\cite{KKI2,KSI}
\begin{eqnarray}  
Z_{\rm qXY}&\equiv& \int \Pi_{\alpha=1,2,3}[d\omega_{\alpha,i}][d\lambda_{i}]
e^{A_{\rm qXY}},\nonumber\\
A_{\rm qXY}&=&A_{\tau}+A_{L}(e^{i\Omega_{\sigma}},e^{-i\Omega_{\sigma}})
+A_{\rm zNN},
\label{Zeff}
\end{eqnarray}
where
\begin{eqnarray}  
A_{\tau}&=&-c_{\tau}\sum_{i}\sum^{3}_{\sigma=1}\cos(\omega_{\sigma,i+{\hat \tau}}-\omega_{\sigma,i}+\lambda_{i}),\\
A_{L}&=&\sum_{\langle i,j\rangle}(C_{1}\cos(\Omega_{1,i}-\Omega_{1,j}) 
+C_{2}\cos(\Omega_{2,i}-\Omega_{3,j})\nonumber\\
&&+C_{3}\cos(\Omega_{2,i}-\Omega_{3,j})),
\end{eqnarray}
and
\begin{eqnarray}  
A_{\rm zNN}&=&-{J}_{\rm zNN}\sum_{\langle i,j\rangle}
\Delta\rho_{i}\Delta\rho_{j}  \nonumber \\
&&-{J}_{\rm zNNN}\sum_{\langle\langle i,l\rangle\rangle}
\Delta\rho_{i}\Delta\rho_{l},
%&&\Delta\rho_{i}\equiv \rho_{1,i}-\rho_{2,i},\nonumber
\end{eqnarray}
where $\langle \cdots\rangle$ stands for the NN sites in the 2D spatial lattice and $\langle\langle \cdots\rangle\rangle$ the NNN ones.
The dynamical variables $\Omega_{\alpha,i}$ ($\alpha=1,2,3$) are related to the phases $\omega_{\alpha,i}$ as
\begin{eqnarray}  
\Omega_{1,i}&=&\omega_{1,i}-\omega_{2,i},\nonumber\\
\Omega_{2,i}&=&\omega_{1,i}-\omega_{3,i},\nonumber\\
\Omega_{3,i}&=&\omega_{2,i}-\omega_{3,i}.\nonumber
\end{eqnarray}

As we explained in Sec.II, the partition function in Eq.(\ref{Zeff}) has been derived by integrating out the amplitude modes of slave-particle fields. 
As a result, the coefficients in the action $A_{\rm qXY}$ depend on the 
local variational parameter $\{\Delta\rho_{i}\}$ and they are explicitly given as
\begin{eqnarray}  
c_{\tau}&=&\frac{1}{V_{0}\Delta\tau},\nonumber\\
C_{1}&=&J_{\rm XY}\rho_{3}\Delta\tau\sqrt{((1-\rho_{3})^{2}-(\Delta\rho_{i})^{2})((1-\rho_{3})^{2}-(\Delta\rho_{j})^{2})},\nonumber\\
C_{2}&=&t\rho_{3}\Delta\tau\sqrt{(1-\rho_{3}+\Delta\rho_{i})(1-\rho_{3}+\Delta\rho_{j})},\nonumber\\
C_{2}&=&t\rho_{3}\Delta\tau\sqrt{(1-\rho_{3}-\Delta\rho_{i})(1-\rho_{3}-\Delta\rho_{j})}.
\nonumber
\end{eqnarray}
By the relation $1/(k_{B}T)=L\cdot \Delta\tau$, where $L$ is the linear system 
size, $\Delta\tau$ has dimension 1/(energy) and the low-temperature
limit is realized for $L\rightarrow\infty$.
We put $c_{\tau}=2$ in the practical calculation, and then
$k_{B}T=(c_{\tau}V_{0})/L=2V_{0}/L$.
Here it should be noticed that a change of the value of $V_0$ results in a 
change of $c_\tau$.
The previous study\cite{KKI2} showed that the global phase structure of the system 
$H_{\rm tJ}$ is stable against to change of the value of $c_\tau$ with fixed $\Delta\tau$.
In general, for smaller value of $V_0$, i.e., a larger $c_\tau$, the parameter 
region of the SF is enlarged\cite{KKI2}.

The partiton function $Z_{\rm qXY}$ in Eq.(\ref{Zeff}) is a functional of 
{$\Delta\rho_{i}$}, i.e., $Z_{\rm qXY}=Z_{\rm qXY}(\{\Delta\rho_{i}\})$.
We expect that {$\Delta\rho_{i}$} behave as variational variables and determine them under the optimal free-energy condition.
In the practical calculation, we performed the local update of {$\Delta\rho_{i}$} by QMC simulation and obtained
\begin{eqnarray}  
[Z_{\rm qXY}]\equiv\int[d\Delta\rho_{i}] Z_{\rm qXY}(\{ \Delta\rho_{i}\}).
\end{eqnarray}
However in the updates of the QMC, $\{\Delta\rho_{i}\}$ are quite 
stable\cite{KSI} for given values of parameters in the action $A_{\rm qXY}$. 
This fact indicates that {$\{\Delta\rho_{i}\}$} should be regarded as variational 
parameters rather than dynamical variables.

For the QMC, we employed the ground-canonical ensemble, 
and therefore the numbers of $a$ and $b$-bosons, $N_{a}$ and $N_{b}$,
are not conserved independently in QMC updates, although 
the total atomic number $N_{a}+N_{b}$ is conserved.

In the practical calculation, we employed the standard Metropolis algorithm with the
local update\cite{Met}.
The typical sweeps for the measurement is (50000-100000)$\times$(10 samples), 
and the acceptance ratio is 40-50 \%.
Errors are estimated from 10 samples by the jackknife method.

To obtain the phase diagram, we calculated the internal energy $E$ and 
specific heat $C$, which are defined as
\begin{eqnarray}  
E=\langle(A_{L}+A_{\rm zNN}) \rangle /L^3,\nonumber\\
C=\langle((A_{L}+A_{\rm zNN})-E)^{2} \rangle /L^3.
\end{eqnarray}
To identify various phases, we also calculated the following pseudo-spin 
correlation function, boson correlation function
and also the density-difference correlation function,
\begin{eqnarray}  
&&G_{S}(r)=
\frac{1}{L^3}\sum_{i_{0}}\langle e^{i\Omega_{1,i_{0}}}e^{-i\Omega_{1,i_{0}+r}}\rangle,\nonumber\\
&&G_{a}(r)=
\frac{1}{L^3}\sum_{i_{0}}\langle e^{i\Omega_{2,i_{0}}}e^{-i\Omega_{2,i_{0}+r}}\rangle,\nonumber\\
&&G_{b}(r)=
\frac{1}{L^3}\sum_{i_{0}}\langle e^{i\Omega_{3,i_{0}}}e^{-i\Omega_{3,i_{0}+r}}\rangle,\nonumber\\
&&G_{dd}(r)=
\frac{1}{L^3}\sum_{i_{0}}\langle \Delta\rho_{i_{0}} \Delta\rho_{i_{0}+r}\rangle,
\label{CRFs}
\end{eqnarray}
where sites $i_{0}$ and $i_{0}+r$ are located in the same spatial 2D lattice. 
The order of the phase transition was identified by calculating the density of 
state $N(E)$ that is defined by
\begin{eqnarray}  
[Z_{\rm qXYZ}]=\int dE N(E) e^{-E}.
\end{eqnarray}
If $N(E)$ has a single peak at the transition point, the phase transition
is of second order.
On the other hand, a double-peak shape of $N(E)$ indicates the
existence of a first-order phase transition.

%%%%%%%%%%%%%%%%%%%%%%%%%%%%%%%%%%%%%%%%%%%%%%%%%%%%%%%%%%%
%FIG.3
\begin{figure}[ht]
\begin{center}
\includegraphics[width=7cm]{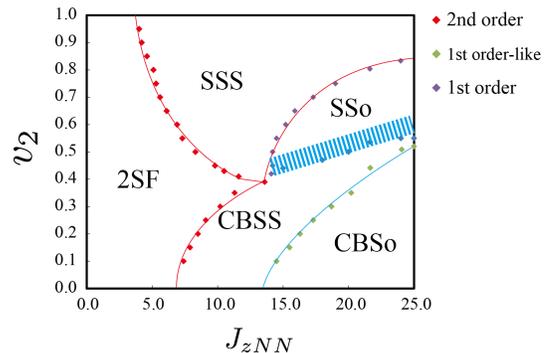}
\caption{(Color online)
Phase diagram at $T=0$ obtained by the QMC,
which include effects of the quantum fluctuations. 
We consider the case with the homogeneous hole density $\rho_{3}=0.3$, 
$c_{\tau}=2.0$, $C_{1}=2$, and $t=20$. 
We introduce the amplitude ratio, $v_{2}=J_{zNNN}/J_{zNN}$.
There exist five phases including two SSs phase, i.e., the CBSS and SSS.
In the blue-shaded region, coexistence of the CBSS and SSo is verified.
%System size $L=16$
}
\label{PD}
\end{center} \vspace{-0.2cm}
\end{figure}
%%%%%%%%%%%%%%%%%%%%%%%%%%%%%%%%%%%%%%%%%%%%%%%%%%%%%%%%%%%% 
%%%%%%%%%%%%%%%%%%%%%%%%%%%%%%%%%%%%%%%%%%%%%%%%%%%%%%%%%%%
%FIG.4
\begin{figure}[t]
\begin{center}
\includegraphics[width=8.3cm]{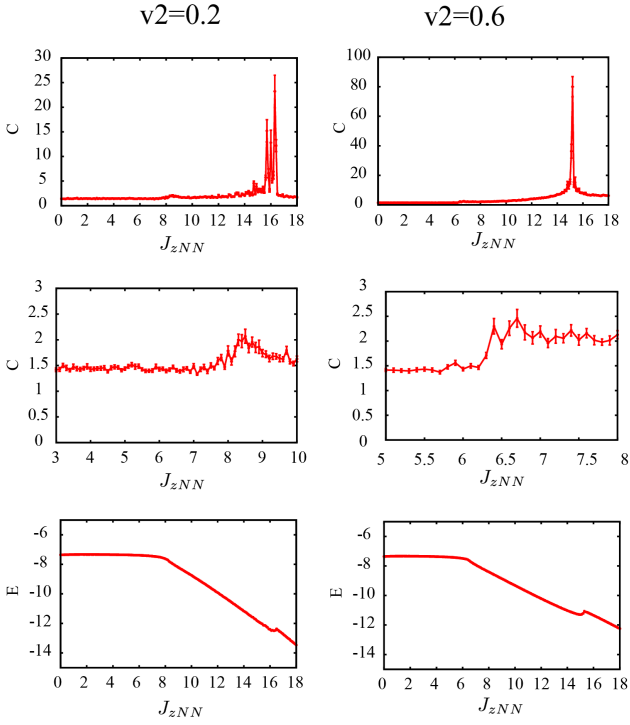}
\caption{(Color online)
The specific heat $C$ and the internal energy $E$ for $v_2=0.2$ and $0.6$.
For both cases, there exist two phase transitions.
See the phase diagram in Fig.\ref{PD}.
System size $L=16$}
\label{C-E-N}
\end{center} \vspace{-0.5cm}
\end{figure}
%%%%%%%%%%%%%%%%%%%%%%%%%%%%%%%%%%%%%%%%%%%%%%%%%%%%%%%%%%%
%%%%%%%%%%%%%%%%%%%%%%%%%%%%%%%%%%%%%%%%%%%%%%%%%%%%%%%%%%%
%FIG.5
\begin{figure}[h]
\begin{center}
\includegraphics[width=8cm]{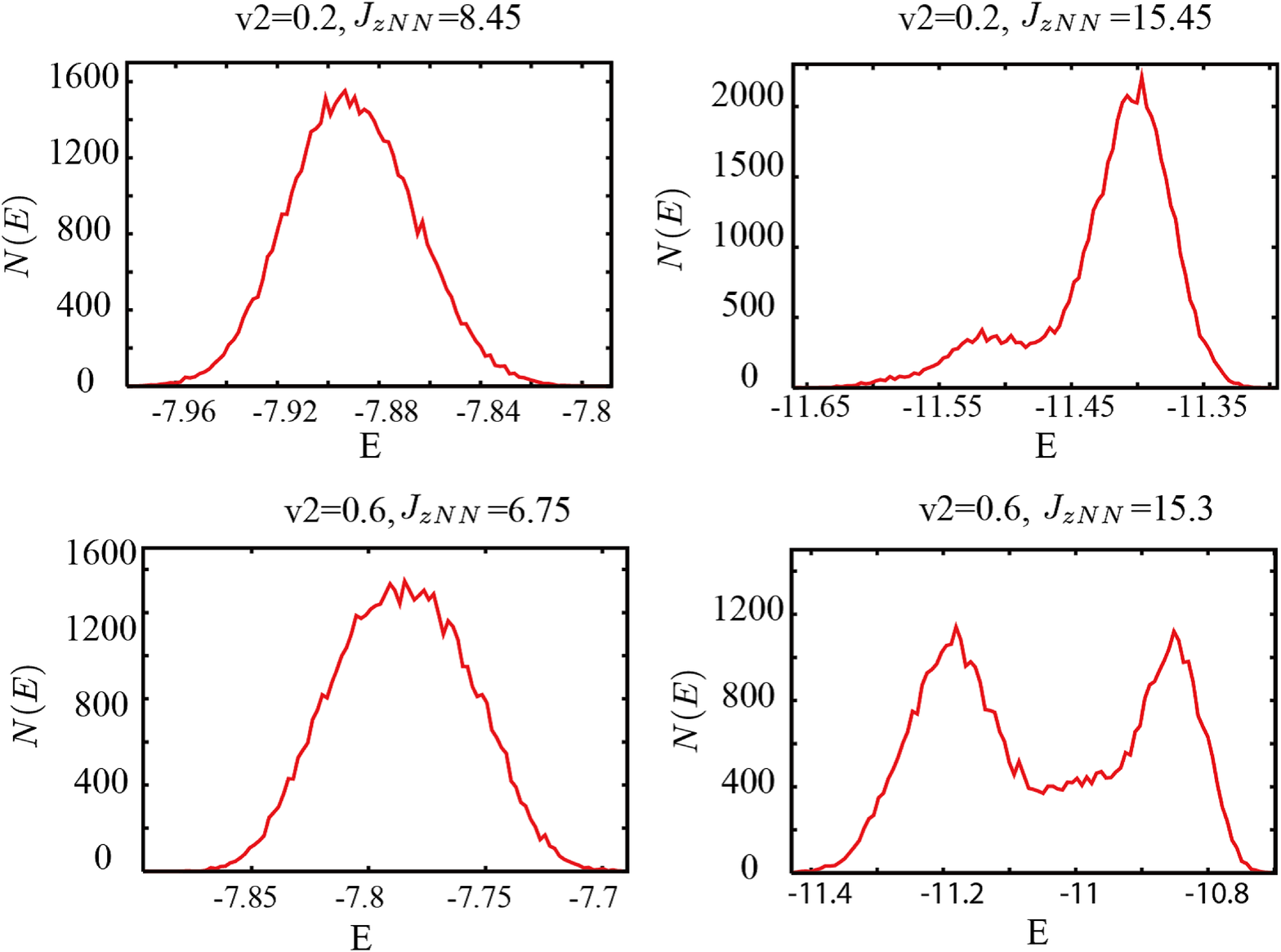}
\caption{(Color online)
Density of state $N(E)$ used to determine order of the phase transitions.
Single peak of $N(E)$ at the phase transition point indicates a second-order
phase transition whereas double peak a first-order one.
$v_{2}=J_{zNNN}/J_{zNN}$, and system size $L=16$}
\label{NE}
\end{center}  \vspace{-0.5cm}
\end{figure}
%%%%%%%%%%%%%%%%%%%%%%%%%%%%%%%%%%%%%%%%%%%%%%%%%%%%%%%%%%%

\begin{figure}[htb]
\begin{center}
\includegraphics[width=7cm]{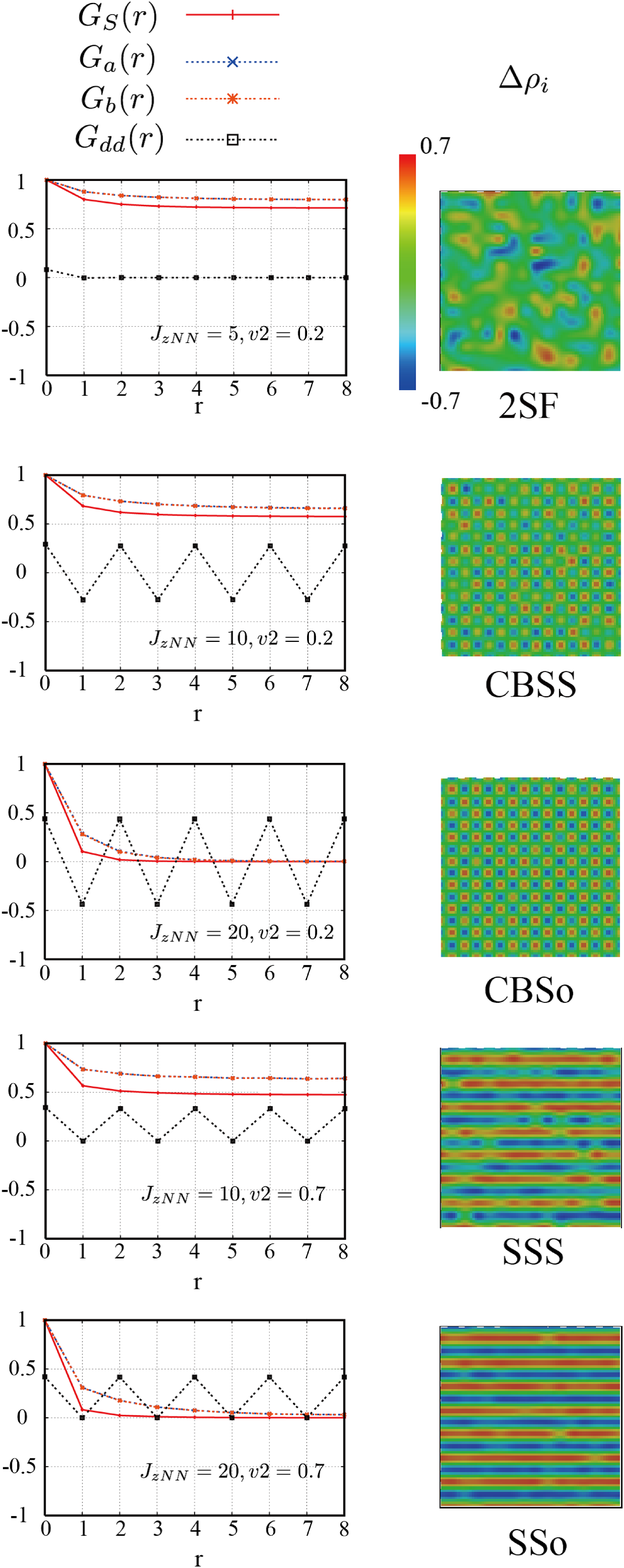}
\caption{(Color online)
Various correlation functions and snapshots used to identify physical 
properties of each phase.
Density difference $\Delta\rho_i\equiv n_{a,i}-n_{b,i}$.
}
\label{Corr}
\end{center}  \vspace{-0.5cm}
\end{figure}

In Fig.\ref{PD}, we show the global phase diagram obtained by the QMC
for $c_{\tau}=2.0$, $C_{1}=2$, and $t=20$.
By calculating the density of states $N(E)$, 
the order of the phase transitions has been determined as indicated in Fig.\ref{PD}.
Typical behaviors of the specific heat $C$ and the internal energy $E$ are 
shown in Fig.\ref{C-E-N} in the $(v_2-J_{\rm zNN})$-plain, where
$v_2=J_{\rm zNNN}/J_{\rm zNN}$.
The density of state, $N(E)$, on typical critical points is shown in Fig.\ref{NE}. 
Furthermore, some correlation functions and density difference $\{\Delta\rho_{i}\}$
snapshots, which were used for the identification of each phase, are 
exhibited in Fig.\ref{Corr}.

%%%%%%%%%%%%%%%%%%%%%%%%%%%%%%%%%%%%%%%%%%%%%%%%%%%%%%%%%%%
%FIG.6
%%%%%%%%%%%%%%%%%%%%%%%%%%%%%%%%%%%%%%%%%%%%%%%%%%%%%%%%%%%
%%%%%%%%%%%%%%%%%%%%%%%%%%%%%%%%%%%%%%%%%%%%%%%%%%%%%%%%%%%
%FIG.7
\begin{figure}[htb]
\begin{center}
\includegraphics[width=7.5cm]{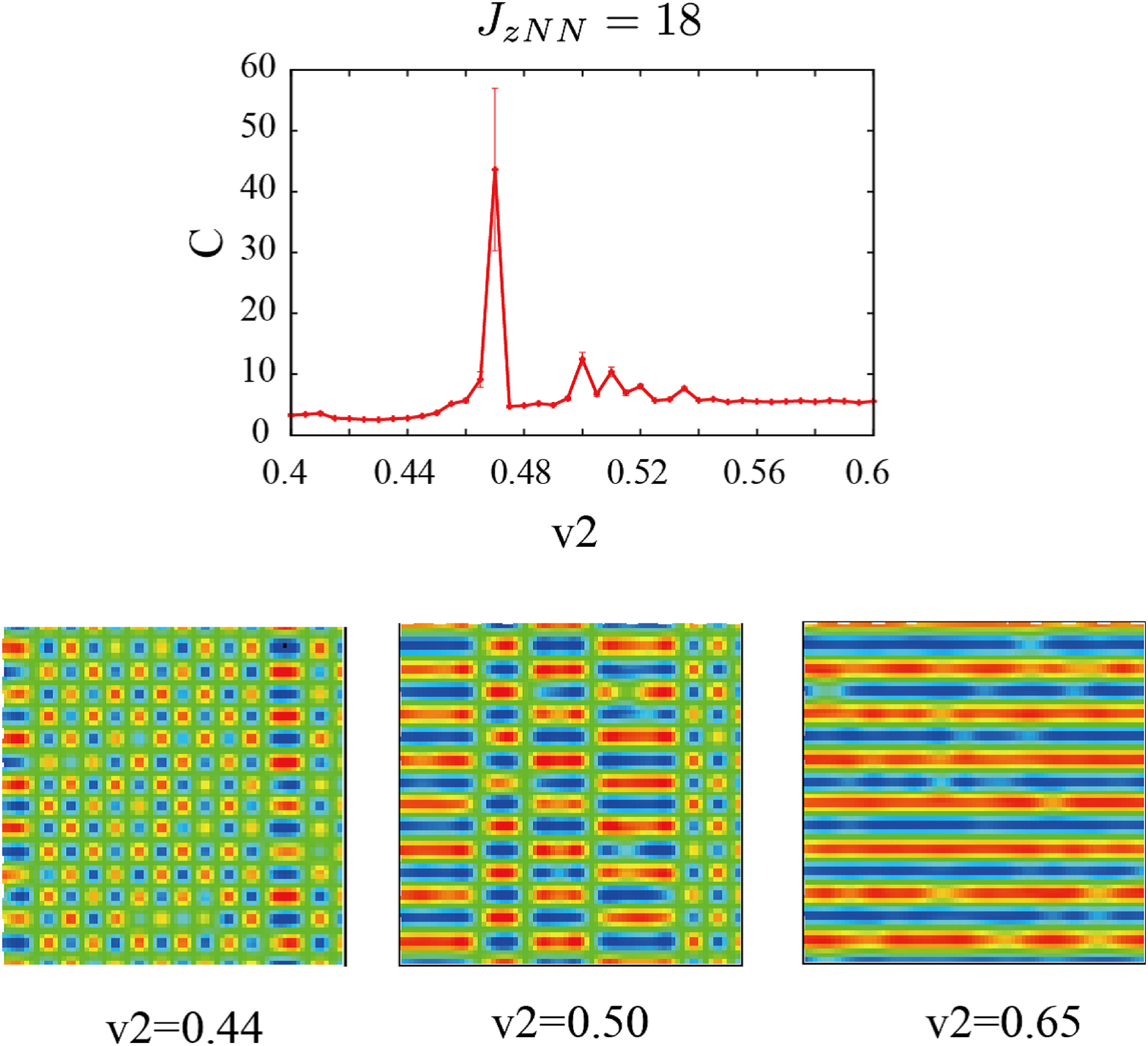}
\caption{(Color online)
Specific heat as a function of $v_2$ for $J_{\rm zNN}=18$.
Besides the large peak at $v_2\simeq 0.47$, which indicates existence of a
first-oder phase transition, there are several small peaks from
$v_2=0.48$ to $0.54$.
Snapshot for $v_2=0.50$ indicates a coexisting phase of the CBSS and 
SSo.
}
\label{Hetero}
\end{center}   \vspace{-0.5cm}
\end{figure}
%%%%%%%%%%%%%%%%%%%%%%%%%%%%%%%%%%%%%%%%%%%%%%%%%%%%%%%%%%%

As the phase diagram in Fig.\ref{PD} shows, there exist five phase: 
2SF, checkerboard supersolid (CBSS), stripe supersolid (SSS), 
CBSo, and SSo. 
In particular, the two kinds of SS form in the 
intermediate parameter regime between the genuine SF and solids. 
The correlation functions indicating the existence of the SS's
are shown in Fig.\ref{Corr}. 

%About MFT
In contrast to the MFT phase diagram of Fig.\ref{PD1}, the SS's form
in rather large parameter region of the phase diagram in Fig.\ref{PD}. 
This means that the quantum fluctuations play an essentially important role
for the coexisting of the SF and solid order.
In other words, in the SS states, both the density of particles and SF
order parameter (i.e., the phase of the boson fields) fluctuate 
as required by the quantum uncertainty principle but their fluctuations are
rather moderate and then the both orders are preserved intact.
It is interesting to notice that the parameter region of the SSS is 
larger than that of the CBSS.
This means that the one-dimensional structure of the stripe is more compatible
with the SF rather than the CB as it is physically expected.
 
%%%%%%%%%%

%About SS existance
%%%%%%%%
As far as the phase diagram in Fig.\ref{PD} shows, there is no direct phase transition
from the CBSo and SSo.
In Ref.\cite{Batrouni2}, a similar phase diagram was reported for the single-component 
Bose Hubbard model.
There the CBSo and SSo are separated by the simple SF phase.
In the present system, however, the CBSS exists between the CBSo and SSo.
%%%%%%%%

%%%%%%%%%%

%Hetero-Structure-like phase transition
%%%%%%%%%%
By the practical calculation,
we have found that some interesting ``phase" exists between the CBSS and SSo, 
which is indicated by the blue-shaded region in the phase diagram in Fig.\ref{PD}.
The specific heat $C$ for $J_{\rm zNN}=18$ has the behavior shown in 
Fig.\ref{Hetero}.
It is obvious that there exists a first-order phase transition at 
$v_2= {{J}_{\rm zNN} \over {J}_{\rm zNNN}}\simeq 0.47$, and 
the CBSo terminates there.
As the value of $v_2$ is increased from $0.47$, several small peaks appear in $C$ 
till $v_2\simeq 0.54$.
Snapshots are quite useful to understand what happens in that region.
See Fig.\ref{Hetero}, in particular, the snapshot of $v_2=0.50$.
Small regions of the CBSo and SSo coexist there in the phase-separated form, 
and we verified that the spatial pattern of these small regions is rather stable 
under the MC updates.
Our observation indicates that there exist several (meta)stable `mixed crystals'
of the CBSo and SSo between the CBSS and SSo, and this mixing of
the solid order destroys the SF.
For the two-dimensional $J_1$-$J_2$ Heisenberg model, it was expected that 
a quantum spin liquid exists between the N\'{e}el state and the stripe
antiferromagnetic state\cite{J1J2}.
The DDI in Eq.(\ref{VDDI}) has a similar structure to the above $J_1$-$J_2$ 
Heisenberg model, but we think the present `mixed crystals' is different
from the quantum liquid as the spatial pattern is stable. 
This is a result of the Ising-type spin coupling of $V_{\rm DDI}$
in contrast to the O(3) symmetric one in the $J_1$-$J_2$ Heisenberg model.

%%%%%%%%%%%%%%%%%%%%%%%%%%%%%%%%%%%%%%%%%%%%%%%%%%%%%%%%%%%%

%\subsection{Effects of artificial external magnetic field on SS${\rm s}$}

Nowadays, it is possible to apply an artificial 
external magnetic field to the atomic system in an OL by rotating
the system or using lasers\cite{art}.
The atomic systems in an artificial magnetic field mimic the superconducting
system, system of the quantum Hall effect, etc., and therefore they are one of
the most interesting subjects in the cold atomic physics.
In this section, we shall study the (in)stability of the SS's in an external
magnetic field.
We expect that the stability depends on the type of the solid order
of the SS's.

%%%%%%%%%%%%%%%%%%%%%%%%%%%%%%%%%%%%%%%%%%%%%%%%%%%%%%%%%%%
%FIG.8
\begin{figure}[h]
\begin{center}
\includegraphics[width=9cm]{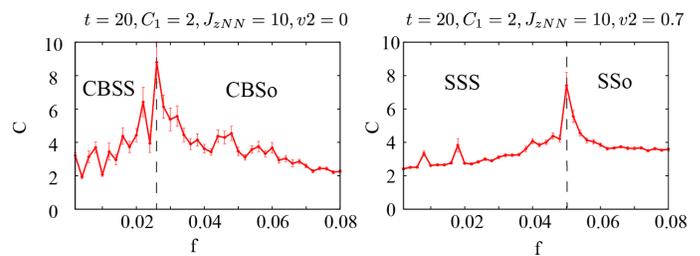}
\caption{(Color online)
Specific heat as a function of the magnetic flux $f$.
As $f$ is increased, SF is lost.
However, the solid orders are stable against the magnetic field.
}
\label{SSsinM} \vspace{-0.5cm}
\end{center}
\end{figure}
%%%%%%%%%%%%%%%%%%%%%%%%%%%%%%%%%%%%%%%%%%%%%%%%%%%%%%%%%%% 

In the practical calculation, we used the symmetric gauge for the
vector potential.
Magnetic flux per plaquette of the OL is denoted by $2\pi f$.
In Fig.\ref{SSsinM}, we show the specific heat as a function of the
strength of the magnetic field $f$.
For the CBSS, the SF is lost at $f\simeq 0.025$, and for the SSS
$f\simeq 0.055$.
This result means that the SSS is more robust than the CBSS
as it is expected from the phase diagram in Fig.\ref{PD}.

%%%%%%%%%%%%%%%%%%%%%%%%%%%%%%%%%%%%%%%%%%%%%%%%%%%%%%%%%%%%

\setcounter{equation}{0}
\section{MC Simulation of $\mbox{t-J}$-like model realized by cold atoms 
in an optical lattice}

In this section we focus on the experiment of A.de Paz et al.\cite{Santos}. 
They succeeded to create the NNN pseudo-spin interactions by using the 
DDI of ${}^{52}$Cr with total spin $s=3$. 
In the experiment, the doubly-occupied states were excluded by the strong on-site
repulsion.
Furthermore by applying an external magnetic field, 
the two states with spin component $m_{s}=-3$ and $m_{s}=-2$ in the direction of 
the magnetic field dominate the system.
Then the reduced DDI is regarded as a pseudo-spin interaction similar to that
of the t-J model. 
The resultant atomic system of ${}^{52}$Cr is a strongly-correlated system and is 
well described by the B-t-J model.

However as shown by the calculation in Ref.\cite{Santos}, the resultant B-t-J model 
has anisotropy in both the hopping amplitudes and pseudo-spin interactions.
Then we call the model {\it t-J-like model} hereafter.
We carried out detailed study of the t-J-like model by the QMC
and obtained the phase structure of the t-J-like model in the parameter regime 
realized in the experiment.  

The effective action Eq.(\ref{Zeff}) changes to the following one by the anisotropy,
\begin{eqnarray}
A_{\rm t-J-like}
&=&-\sum_{\tau,i}c_{\tau}\cos(\theta_{a,i}-\theta_{a,i+\tau})
+c_{\tau}\cos(\theta_{b,i}-\theta_{b,i+\tau})\nonumber\\
&+&\sum_{i,j\in NNN}C_{xy,j}\cos((\theta_{a,i}-\theta_{b,i})-(\theta_{a,j}-\theta_{b,j}))
\nonumber\\
&-&\sum_{i,j\in NN}C_{2,j}(\cos(\theta_{a,i}-\theta_{a,j})+\cos(\theta_{b,i}-\theta_{b,j}))\nonumber\\
&+&\sum_{i,j\in NNN}C_{z,j}\Delta\rho_{i}\Delta\rho_{j}, %\nonumber
\label{AtJL}
\end{eqnarray}
where
\begin{eqnarray}
%&&c_{\tau}=\frac{1}{\Delta\tau V_{0}},\nonumber\\
&&C_{xy,j}=-\frac{1}{2}V_{ij}\frac{1}{4}\Delta\tau\sqrt{(\rho^{2}_{0}
-\Delta\rho^{2}_{i})(\rho^{2}_{0}-\Delta\rho^{2}_{j})},\nonumber\\
&&C_{2,j}=t_{j}\frac{1}{2}\Delta\tau\sqrt{(\rho_{0}-\Delta\rho_{i})
(\rho_{0}-\Delta\rho_{j})},\nonumber\\
&&C_{z,j}=V_{ij}\Delta\tau, %\nonumber
\label{Cs2}
\end{eqnarray}
and the anisotropic couplings $t_{j}$ and $V_{i,j}$ are given as 
\begin{eqnarray}
V_{i,j}=\left\{ \begin{array}{ll}
0.8W_{0} & (j=i+\hat{x},i-\hat{x}), \\
-1.8W_{0} & (j=i+\hat{y},i-\hat{y}), \\
-0.11W_{0} & (j\in {\rm NNN}), \\
\end{array} \right.   %\nonumber
\label{VijW0}
\end{eqnarray}
\begin{eqnarray}
t_{j}=\left\{ \begin{array}{ll}
3.66t & (j=i+\hat{x},i-\hat{x}), \\
t & (j=i+\hat{y},i-\hat{y}), \\
\end{array} \right.  %\nonumber
\label{tjaniso}
\end{eqnarray}
with
\begin{eqnarray}
W_{0}=\frac{\mu_{0}\mu^{2}_{B}}{\pi(a_L /2)^{3}},   %\nonumber
\label{W0}
\end{eqnarray}
where $\mu_{B}$ is the Bohr magneton and $\mu_0$ is the magnetic permeability
of vacuum as before.
The proposed t-J-like model in Ref.\cite{Santos} has an additional effective Zeeman 
coupling along $S^z$, but we ignore it in the present study because we are 
interested in the genuine effect of the DDI.
    
For the practical calculation, we regard the $W_{0}$ as a free parameter and put 
the hole density $\rho_{0}=0.3$.
The strength of the dipole-induced pseudo-spin interaction relative to the hopping
amplitude $t$ determines the equilibrium state.

In the experiment, it was observed that there exists a density difference between
the $m_s=-3$ and $m_s=-2$ states in the equilibrium, 
and this phenomenon was considered as a result of the DDI.

%%%%%%%%%%%%%%%%%%%%%%%%%%%%%%%%%%%%%%%%%%%%%%%%%%%%%%%%%%%
%FIG.9
\begin{figure}[h]
\begin{center}
\includegraphics[width=9cm]{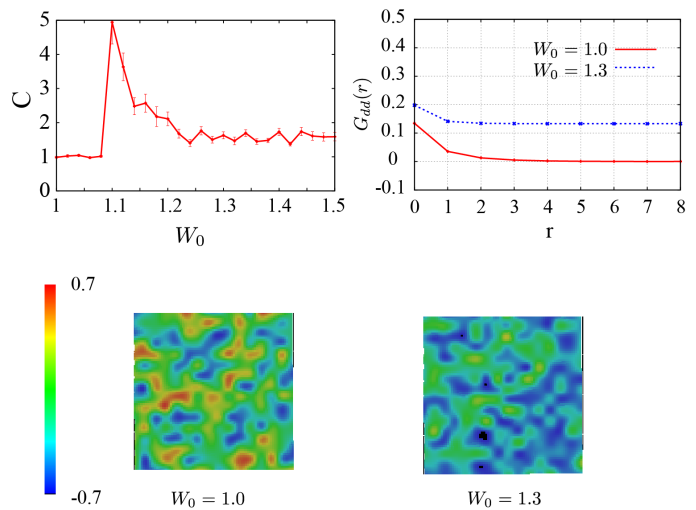}
\caption{(Color online)
Up-left panel: The specific heat with $t=2$, $c_{\tau}=2$. System size $L=16$.
Up-right panel: Two typical behaviors of the density difference correlation. 
Each of two states is equilibrium.
Bottom-left:Snapshot of solid order (the density difference) for $W_0=1.0$.
Bottom-right:Snapshot of solid order for $W_0=1.3$
It seems that the both cases have no clear solid order, but we verified that 
the density pattern is quite stable for the MC update.
}
\label{C_t-J-like}
\end{center}
\end{figure}
%%%%%%%%%%%%%%%%%%%%%%%%%%%%%%%%%%%%%%%%%%%%%%%%%%%%%%%%%%% 

In Fig.{\ref{C_t-J-like}}, we show the behaviors of the specific heat $C$ for 
the t-J-like model with the energy unit $t=2$.
The obtained specific heat exhibits the existence of a second-order phase transition 
at $W_{0c}\simeq 1.1$.
We also calculated the density-difference correlation function (DDCF) 
$G_{dd}(r)$ defined by Eq.(\ref{CRFs}).
From the DDCF shown in Fig.\ref{C_t-J-like}, it is obvious that 
$G_{dd}(r) \rightarrow \mbox{finite} \ (0)$ as $r\rightarrow \mbox{large}$
for $W_0>W_{0c} \ (W_0<W_{0c})$, i.e.,
the density of one atom is globally larger
than that of the other for $W_0>W_{0c}$, whereas
the equal density distribution is realized for $W_0<W_{0c}$.
This is the direct result of the dipolar intersite spin interaction,
and is in agreement with the experimental observation.
Density snapshots for $W_0>W_{0c}$ and $W_0<W_{0c}$ are shown in Fig.\ref{C_t-J-like}.
No specific spatial pattern is observed in contrast to the previous case
that is studied in Sec.IV.

%\vspace{1cm}   
%%%%%%%%%%%%%%%%%%%%%%%%%%%%
\section{Conclusion}
\setcounter{equation}{0}

In this paper, we studied the extended B-t-J model of the two-component 
bosons with the long-range DDI.
We show the DDI can generate the additional pseudo-spin interactions
by controlling directions of the dipoles of the $a$ and $b$-atoms.
We studied the global phase diagram of the extended B-t-model by means of the
Gutzwiller variational method and the QMC.
Obtained phase diagrams indicate that quantum fluctuation is an essential 
ingredient for the realization of the SS's.
The QMC predicts two kinds of the SS state, one of which is the CBSS and
the other is the SSS, and the latter stems from the long-range nature of the DDI.
Detailed study of the phase boundary of the CBSS and SSo was also given.

Finally we investigated the t-J-like model, which is expected to describe the 
strongly-correlated system recently realized by the experiment\cite{Santos}. 
By the QMC, we confirmed the existence of the phase transition as the strength 
of the DDI is increased.
In the state with the DDI stronger than the critical one, an imbalance of the
density of atoms, which is nothing but a finite pseudo-spin order in the $z$-direction,
appears. 
The obtained results are consistent with the experimental findings.

%%%%%%%%%%%%%%%%%%%%%%%%%%%%%%%%%%%%%%%%%%%%%%%%%%%
%\bigskip 
\acknowledgments 
This work was partially supported by Grant-in-Aid
for Scientific Research from Japan Society for the 
Promotion of Science under Grant No.26400246.
%%%%%%%%%%%%%%%%%%%%%%%%%%%%%%%%%%%%%%%%%%%%%%%%%%

%%%%%%%%%%%%%%%%%%%%%%%%%%%%


\begin{references}
\bibitem{HCboson}It is rather straightforward to treat $a_i$ and $b_i$ as 
hard-core bosons as in Refs.\cite{BtJ,BtJ1}.
As we impose the local constraint excluding the doubly-occupied state
in the following calculation, obtained results are qualitatively the same.
\end{references}
\end{document}